\documentclass[preprint,aps,12pt,showpacs,tightenlines]{revtex4}
\usepackage{amsmath}
\usepackage{amssymb}
\usepackage{epsfig}
\usepackage{graphicx}
\begin{document}
\preprint{NJNU-TH-06-24}
%%%%%%%%%%%%%%%%%%%%%%%%%%%%%%%%%%%%%%%%%%%%%

\newcommand{\beq}{\begin{eqnarray}}
\newcommand{\eeq}{\end{eqnarray}}
\newcommand{\non}{\nonumber\\ }

\newcommand{\acp}{ {\cal A}_{CP} }
\newcommand{\psl}{ p \hspace{-1.8truemm}/ }
\newcommand{\nsl}{ n \hspace{-2.2truemm}/ }
\newcommand{\vsl}{ v \hspace{-2.2truemm}/ }
\newcommand{\epsl}{\epsilon \hspace{-1.8truemm}/\,  }

\def \cpl{ Chin. Phys. Lett.  }
\def \ctp{ Commun. Theor. Phys.  }
\def \epjc{ Eur. Phys. J. C }
\def \jpg{  J. Phys. G }
\def \npb{  Nucl. Phys. B }
\def \plb{  Phys. Lett. B }
\def \prd{  Phys. Rev. D }
\def \prl{  Phys. Rev. Lett.  }
\def \zpc{  Z. Phys. C }
\def \jhep{ J. High Energy Phys.  }

%%%%%%%%%%%%%%%%%%%%%%%%%%%%%%%%%%%%%%%%%%%%%%%%%%%%
%%
\title{Branching Ratio and CP Asymmetry of $B_{s} \to \rho(\omega) K $ Decays in the
Perturbative QCD Approach}
\author{Zhen-jun Xiao \email{xiaozhenjun@njnu.edu.cn}, Xin-fen Chen and Dong-qin Guo}
\affiliation{Department of Physics and Institute of Theoretical Physics, Nanjing Normal
University, Nanjing, Jiangsu 210097, P.R.China}
\date{\today}
\begin{abstract}
In this paper, we calculate the branching ratios and CP-violating
asymmetries for $B_{s}\to \rho^\pm K^\mp$,
$B_{s}\to\rho^{0}\overline{K}^0 $ and
$B_{s}\to\omega\overline{K}^0$ decays in
the perturbative QCD (pQCD) factorization approach.
The theoretical predictions for the CP averaged branching
ratios of the considered decays are: $Br(B_{s}\to\rho^\pm K^\mp )\approx 24.7 \times
10^{-6}$, $Br(B_{s}\to\rho^{0}\overline{K}^0 ) \approx 1.2 \times 10^{-7}$
and $Br(B_{s} \to\omega\overline{K}^0 )
\approx 1.7 \times 10^{-7}$; and  we also predict very large direct
CP-violating asymmetries for  the latter two decay modes:
$A_{CP}^{dir}(\rho^{\pm} K^{\mp})\approx -12 \%$,
$A_{CP}^{dir}(\rho^0\overline{K}^0)\approx -92\%$,
$A_{CP}^{dir}(\omega^0\overline{K}^0)\approx 81\%$,
$A_{CP}^{mix}(\rho^0\overline{K}^0)\approx -36\%$, and
$A_{CP}^{mix}(\omega^0\overline{K}^0) \approx -40\%$,
which can be tested in the forthcoming LHC-b experiments.
\end{abstract}


\maketitle

\section{Introduction}

The charmless B meson decay is a good place to test the Standard Model (SM), study CP
violation and look for signal or evidence of possible new physics beyond the SM.
Since 1999, many such decay modes have been observed in the B factory experiments.
In the forthcoming Large Hadron Collider beauty (LHC-b) experiments,
a large number of heavier $B_s$ and $B_c$ mesons together with light $B_{u,d}$ mesons will  be produced
\cite{lhcb1}.
The study about the charmless decays of $B_s$ meson therefore is therefore
becoming more interesting then ever before \cite{lhcb2}.

By employing the generalized facterization approach\cite{zpc,aag,yhc} or the QCD factorization (QCDF)
approach \cite{bbns99},
about 40 $B_s \to h_1 h_2$ ($h_i$ stand for light pseudo-scalar or vector mesons ) decay modes
have been studied in the framework of SM \cite{chenbs99,tseng99,sunbs03,mb333} or in some new physics models
beyond the SM \cite{xiaobs01}. In Refs.~\cite{pipi,pirho,yu-7107,zhu05,pieta}, the branching ratios
and CP violating asymmetries of $B_s\to \pi^{+}\pi^{-}, \pi \rho$, $\pi K$, $\rho(\omega) K^*$
and $\pi \eta^{(\prime)}$ decays have been calculated by employing the perturbative QCD (pQCD)
factorization approach \cite{chc,yyk01}.
Motivated by the expected successes in LHC-b experiments and other hadronic B meson experiments,
we here continue to investigate more charmless $B_s$ decays in pQCD factorization approach.

In this paper, we will study the $B_{s}\to\rho^{\pm} K^{\mp}$ , $\rho^{0}\overline{K}^0$
and $\omega\overline{K}^0$ decays in the pQCD approach.
In principle, the physics for the $B_{s}$ two-body hadronic
decays is very similar to that for the $B_{d}$ meson except that the
spectator $\emph{d}$ quark is replaced by the $\emph{s}$  quark.
Theoretically, the three decays have been studied before in the
naive or generalized factorization approach, as well as in the QCD
factorization approach\cite{sunbs03,mb333}:

For $B_{s} \to \rho(\omega) K $ decays, the $B_{s}$ meson is heavy, setting at rest and decaying
into two light mesons (i.e. $\rho(\omega)$ and $K$ ) with large momenta. Therefore the light
final state mesons are moving very fast in the rest frame of $B_{s}$
meson. In this case, the short distance hard process dominates the
decay amplitude. We assume that the soft final state interaction is
not important for such decays, since there is not enough time for
light mesons to exchange soft gluons. Therefore, it makes the pQCD
reliable in calculating the $B_{s} \to \rho(\omega) K$ decays. With
the Sudakov resummation, we can include the leading double
logarithms for all loop diagrams, in association with the soft
contribution. Unlike the usual factorization approach, the hard part
of the pQCD approach consists of six quarks rather than four. We
thus call it six-quark operators or six-quark effective theory.
Applying the six-quark effective theory to $B_{s}$ meson decays, we
need meson wave functions for the hadronization of quarks into
mesons. All the collinear dynamics are included in the meson wave
functions.

This paper is organized as follows. In Sec.~\ref{sec:f-work}, we
give a brief review for the pQCD factorization approach. In
Sec.~\ref{sec:p-c}, we calculate analytically the related Feynman
diagrams and present the  decay amplitudes for the studied decay
modes. In Sec.~\ref{sec:n-d}, we show the numerical results for the
branching ratios and CP asymmetries of $B_{s} \to \rho(\omega) K$
decays and comparing them with the results obtained in the QCDF approach.
The summary and some discussions are included in the final section.

\section{ Theoretical framework}\label{sec:f-work}

The three scale pQCD factorization approach has been developed and
applied in the non-leptonic $B_{(s)}$ meson decays for some time. In
this approach, the decay amplitude is separated into soft $(\Phi)$,
hard(H), and harder(C) dynamics characterized by different energy
scales $(t, m_b, M_W)$ . It is conceptually written as the
convolution:
 \beq {\cal
 A}(B_{(s)} \to M_1 M_2)\sim \int\!\! d^4k_1 d^4k_2 d^4k_3\
 \mathrm{Tr} \left [ C(t) \Phi_{B_{(s)}}(k_1) \Phi_{M_1}(k_2)
 \Phi_{M_2}(k_3) H(k_1,k_2,k_3, t) \right ], \label{eq:con1}
 \eeq
where $k_i$'s are momenta of light quarks included in each mesons,
and $\mathrm{Tr}$ denotes the trace over Dirac and color indices.
The harder dynamic involves the four quark operators described by
the Wilson coefficient $C(t)$. It results from the radiative
corrections to the four quark operators  at short distance. In the
above convolution, $C(t)$ includes the harder dynamics at larger
scale than $M_{B_{(s)}}$ scale and describes the evolution of local
$4$-Fermi operators from $m_W$ (the $W$ boson mass) down to
$t\sim\mathcal{O}(\sqrt{\bar{\Lambda} M_{B_{(s)}}})$ scale, where
$\bar{\Lambda}\equiv M_{B_{(s)}} -m_b$. The function
$H(k_1,k_2,k_3,t)$ describes the four quark operator and the
spectator quark connected by
 a hard gluon whose scale is at the order
of $M_{B_{(s)}}$,so this hard part $H$ can be perturbatively
calculated.The hard and harder dynamics together make an effective
six-quark interaction. The soft dynamic is factorized into the meson
wave function $\Phi_M$ ,which describes hadronization of the quark
and anti-quark pair into the meson $M$. While the function $H$
depends on the processes considered, the wave function $\Phi_M$ is
independent of the specific processes. Using the wave functions
determined from other well measured processes, one can make
quantitative predictions here.

 For the $B_{s}$ meson decays, since the b quark is
rather heavy we consider the $B_{s}$ meson at rest for simplicity.
It is convenient to use light-cone coordinate $(p^+, p^-, {\bf p}_T)$ to describe the
meson's momenta. The $B_{s}$ meson and the two final state meson momenta can be written as
\beq
P_{B_{s}} = \frac{M_{B_{s}}}{\sqrt{2}} (1,1,{\bf 0}_T), \quad
P_{\rho(\omega)} =
\frac{M_{B_{s}}}{\sqrt{2}}(1,r_{\rho(\omega)}^{2},{\bf 0}_T), \quad
P_{k} = \frac{M_{B_{s}}}{\sqrt{2}} (0,1-r_{\rho(\omega)}^{2},{\bf 0}_T),
\eeq
respectively, where $r_\rho=m_\rho/M_{B_{s}}$; the light
pseudoscalar meson masses have been neglected.

For the $B_{s} \to \rho(\omega) K$ decays considered here, only the
$\rho(\omega)$ meson's longitudinal part contributes to the decays,
its polar vector is
$\epsilon_L=\frac{M_{B_{s}}}{\sqrt{2}M_{\rho(\omega)}}
(1,-r_{\rho(\omega)}^{2},\bf{0_T})$. Putting the light (anti-) quark
momenta in $B_{s}$, $\rho(\omega)$ and $k$ mesons as $k_1$, $k_2$,
and $k_3$, respectively, we can choose
    \beq
    k_1 = (x_1 P_1^+,0,{\bf
    k}_{1T}), \quad k_2 = (x_2 P_2^+,0,{\bf k}_{2T}), \quad k_3 = (0,
    x_3 P_3^-,{\bf k}_{3T}).
     \eeq
  Unlike the QCD factorization approach,we don't neglect the transverse  momentum
  $k_{T}$ in the above expressions, by which to avoid the endpoint
  singularity.Then, the integration over $k_1^-$,
     $k_2^-$, and $k_3^+$ in eq.(\ref{eq:con1}) will lead to
     \beq {\cal
      A}(B_{s} & \to & \rho(\omega) k) \sim \int\!\! d x_1 d x_2 d x_3 b_1 d
      b_1 b_2 d b_2 b_3 d b_3 \non && \quad \mathrm{Tr} \left [ C(t)
      \Phi_{B_{s}}(x_1,b_1) \Phi_{\rho(\omega)}(x_2,b_2) \Phi_{k}(x_3,
      b_3) H(x_i, b_i, t) S_t(x_i)\, e^{-S(t)} \right ], \label{eq:a2}
      \eeq
      where $b_i$ is the conjugate space coordinate of $k_{iT}$, and
$t$ is the largest energy scale in function $H(x_i,b_i,t)$, as a
function in terms of $x_i$ and $b_i$. The large logarithms ($\ln
m_W/t$) coming from QCD radiative corrections to four quark
operators are included in the Wilson coefficients $C(t)$. The large
double logarithms ($\ln^2 x_i$) on the longitudinal direction are
summed by the threshold resummation ~\cite{hnl10}, and they lead to
$S_t(x_i)$ which smears the end-point singularities on $x_i$. The
last term, $e^{-S(t)}$, is the Sudakov form factor resulting from
overlap of soft and collinear divergences, which suppresses the soft
dynamics effectively ~\cite{hnl43}. Thus it makes the perturbative
calculation of the hard part $H$ applicable at intermediate scale,
i.e., $M_{B_{s}}$ scale. We will calculate analytically the function
$H(x_i,b_i,t)$ for $B_{s} \to \rho(\omega) K$ decays in the first
order in $\alpha_s$ expansion and give the convoluted amplitudes in
next section.

\subsection{ Wilson Coefficients}\label{ssec:w-c}

It is well known that the low-energy effective Hamiltonian is the
basic tool to calculate the branching ratios and $\acp$ of
$B$ meson decays. For $B_{s} \to \rho(\omega) K$ decays, the related
weak effective Hamiltonian $H_{eff}$ can be written as
\cite{buras96}
\beq
\label{eq:heff}
{\cal H}_{eff} = \frac{G_{F}} {\sqrt{2}} \, \left[ V_{ub} V_{ud}^* \left (C_1(\mu) O_1^u(\mu) +
C_2(\mu) O_2^u(\mu) \right) - V_{tb} V_{td}^* \, \sum_{i=3}^{10}
C_{i}(\mu) \,O_i(\mu) \right] \; .
\eeq
We specify below the operators in ${\cal H}_{eff}$ for $b \to d$ transition:
\beq
\begin{array}{llllll}
O_1^{u} & = &  \bar d_\alpha\gamma^\mu L u_\beta\cdot \bar
u_\beta\gamma_\mu L b_\alpha\ , &O_2^{u} & = &\bar
d_\alpha\gamma^\mu L u_\alpha\cdot \bar
u_\beta\gamma_\mu L b_\beta\ , \\
O_3 & = & \bar d_\alpha\gamma^\mu L b_\alpha\cdot \sum_{q'}\bar
 q_\beta'\gamma_\mu L q_\beta'\ ,   &
O_4 & = & \bar d_\alpha\gamma^\mu L b_\beta\cdot \sum_{q'}\bar
q_\beta'\gamma_\mu L q_\alpha'\ , \\
O_5 & = & \bar d_\alpha\gamma^\mu L b_\alpha\cdot \sum_{q'}\bar
q_\beta'\gamma_\mu R q_\beta'\ ,   & O_6 & = & \bar
d_\alpha\gamma^\mu L b_\beta\cdot \sum_{q'}\bar
q_\beta'\gamma_\mu R q_\alpha'\ , \\
O_7 & = & \frac{3}{2}\bar d_\alpha\gamma^\mu L b_\alpha\cdot
\sum_{q'}e_{q'}\bar q_\beta'\gamma_\mu R q_\beta'\ ,   & O_8 & = &
\frac{3}{2}\bar d_\alpha\gamma^\mu L b_\beta\cdot
\sum_{q'}e_{q'}\bar q_\beta'\gamma_\mu R q_\alpha'\ , \\
O_9 & = & \frac{3}{2}\bar d_\alpha\gamma^\mu L b_\alpha\cdot
\sum_{q'}e_{q'}\bar q_\beta'\gamma_\mu L q_\beta'\ ,   & O_{10} &
= & \frac{3}{2}\bar d_\alpha\gamma^\mu L b_\beta\cdot
\sum_{q'}e_{q'}\bar q_\beta'\gamma_\mu L q_\alpha'\ ,
\label{eq:operators}
\end{array}
\eeq
where $\alpha$ and $\beta$ are the $SU(3)$ color indices; $L$
and $R$ are the left- and right-handed projection operators with
$L=(1 - \gamma_5)$, $R= (1 + \gamma_5)$. The sum over $q'$ runs over
the quark fields that are active at the scale $\mu=O(m_b)$, i.e.,
$(q'\epsilon\{u,d,s,c,b\})$. The pQCD approach works well for the
leading twist approximation and leading double logarithm summation.
For the Wilson coefficients $C_i(\mu)$ ($i=1,\ldots,10$), we will
also use the leading order (LO) expressions, although the
next-to-leading order   calculations already exist in the literature
~\cite{buras96}. This is the consistent way to cancel the explicit
$\mu$ dependence in the theoretical formulae.

For the renormalization group evolution of the Wilson coefficients
 from higher scale to lower scale, we use the formulae as given in
Ref.~\cite{cdl09} directly. At the high $m_W$ scale, the leading
order Wilson coefficients $C_i(M_W)$ are simple and can be found
easily in Ref.~\cite{buras96}. In pQCD approach, the scale t may be
larger or smaller than the $m_b$ scale. For the case of $ m_b< t<
m_W$, we evaluate the Wilson coefficients at $t$ scale using leading
logarithm running equations, as given in Eq.(C1) of Ref.~\cite{cdl09}.
For the case of $t < m_b$, we then evaluate the Wilson
coefficients at $t$ scale by using $C_i(m_b)$ as input
and the formulae given in Appendix D of Ref.~\cite{cdl09}.

\subsection{Wave Functions}\label{ssec:w-f}

In the resummation procedures, the $B_{s}$ meson is treated as a
heavy-light system. In general, the $B_{s}$meson light-cone matrix
element can be decomposed as\cite{bbns99,agg72}
  \beq
&&\int_0^1\frac{d^4z}{(2\pi)^4}e^{i\bf{k_1}\cdot z}
   \langle 0|\bar{b}_\alpha(0)s_\beta(z)|B(p_{B_{s}})\rangle \nonumber\\
&=&-\frac{i}{\sqrt{2N_c}}\left\{(\psl_{B_{s}}+M_{B_{s}})\gamma_5
\left[\phi_{B_{s}} ({\bf k_1})-\frac{\nsl-\vsl}{\sqrt{2}}
\bar{\phi}_{B_{s}}({\bf k_1})\right]\right\}_{\beta\alpha},
\label{aa1}
 \eeq
 where $n=(1,0,{\bf 0_T})$, and $v=(0,1,{\bf 0_T})$ are the
unit vectors pointing to the plus and minus directions,
respectively. From the above equation, one can see that there are
two Lorentz structures in the $B_{s}$ meson distribution amplitudes.
They obey to the following normalization conditions
 \beq
 \int\frac{d^4 k_1}{(2\pi)^4}\phi_{B_{s}}({\bf
k_1})=\frac{f_{B_{s}}}{2\sqrt{2N_c}}, ~~~\int \frac{d^4
k_1}{(2\pi)^4}\bar{\phi}_{B_{s}}({\bf k_1})=0.
 \eeq

In general, one should consider these two Lorentz structures in
calculations of $B_{s}$ meson decays. However, it can be argued that
the contribution of $\bar{\phi}_{B_{s}}$ is numerically small, thus
its contribution can be numerically neglected. Using this
approximation, we can reduce one input parameter in our calculation.
 Therefore, we only consider the contribution of Lorentz
structure
\beq
\Phi_{B_{s}}= \frac{1}{\sqrt{2N_c}} (\psl_{B_{s}}
+M_{B_{s}}) \gamma_5 \phi_{B_{s}} ({\bf k_1}). \label{bmeson}
\eeq
   In the next section, we
will see that the hard part is always independent of one of the
$k_1^+$ and/or $k_1^-$, if we make  approximations shown in next
section. The $B_{s}$ meson wave function is then the function of
variable $k_1^-$ (or $k_1^+$) and $k_1^\perp$,
 \beq
 \phi_{B_{s}} (k_1^-,
k_1^\perp)=\int d k_1^+ \phi (k_1^+, k_1^-, k_1^\perp). \label{int}
\eeq

The wave function for the pseudoscalar meson K are given as:
\beq
\Phi_{K}(P,x,\zeta)\equiv \frac{i}{\sqrt{2N_c}}\gamma_{5}
\left\{\psl_{K} \phi_{K}^{A}(x)+m_0^{K} \phi_{K}^{P}(x)+\zeta
m_0^{K} (\vsl \nsl-v\cdot n)\phi_{K}^{T}(x)\right\}
\eeq
where $P$ and $x$ are the momentum and the momentum fraction of $K$ meson, respectively.
We assumed here that the wave function of $K$ meson is the same as the wave function of $\pi$
meson. The parameter $\zeta$ is either $+1$ or $-1$
depending on the assignment of the momentum fraction $x$.

For $B \to \rho K$ decay, the $\rho$ meson is longitudinally polarized.
We only consider its wave function in longitudinal polarization ~\cite{tk07,pb23},
 \beq
<\rho^{-}(P,\epsilon_L)|\bar{d_{\alpha}}(z)u_{\beta}(0)|0>=
\frac{1}{\sqrt{2N_c}}\int_0^1 d x e^{ixP\cdot z} \left\{ \epsl
\left[\psl_\rho \phi_\rho^t (x) + m_\rho \phi_\rho (x) \right]
+m_\rho \phi_\rho^s (x)\right\}. \label{eq:brhok}
\eeq
The second term in above
equation is the leading twist wave function (twist-2), while the
first and third terms are sub-leading twist (twist-3) wave
functions. For $B \to \omega K$ decay, we have similar expression as Eq.~(\ref{eq:brhok}).

The transverse momentum $k^\perp$ is usually conveniently
converted to the $b$ parameter by Fourier transformation.
 The initial conditions of function $\phi_i(x)$,
$i=B_{s},\rho,\omega, k$, are of non-perturbative origin, satisfying
the normalization relation
\beq
\int_0^1\phi_i(x,b=0)dx=\frac{1}{2\sqrt{6}}{f_i}\;, \label{no}
\eeq
with $f_i$ the meson decay constants.

\section{Perturbative Calculations}\label{sec:p-c}

In the previous section we have discussed the wave functions and
Wilson coefficients of the amplitude in eq.(\ref{eq:con1}). In this
section, we will calculate the hard part $H(t)$. This part involves
the four quark operators and the necessary hard gluon connecting the
four quark operator and the spectator quark.  We will show the whole
amplitude for each diagram including wave functions. Similar to the
$B_{s} \to \pi K$ ~\cite{yu-7107}, the total 8 lowest order diagrams
contributing to the $B_{s} \to \rho K$ decays, are illustrated in
Figure 1. We first calculate the usual factorizable diagrams (a) and
(b). Operators $O_1$, $O_2$, $O_3$, $O_4$, $O_9$, and $O_{10}$ are
$(V-A)(V-A)$ currents, the sum of their amplitudes is given as
 \beq
 F_{eK}&=& - 8\pi C_F f_\rho
 M_{B_{s}}^{4} \int_0^1 d x_{1} dx_{3}\, \int_{0}^{\infty} b_1 db_1
 b_3 db_3\, \phi_{B_{s}}(x_1,b_1)
 \non
 & & \cdot \left\{ \left [(2-x_3) \phi_{K}^{A} (x_3, b_3)
 +(2x_3-1)r_{K} (\phi_{K}^{P} (x_3, b_3)-\phi_{K}^{T} (x_3,b_3))
 \right] \right.
 \non
 && \left. \alpha_s(t_e^1)
 h_e(x_1,1-x_3,b_1,b_3)\exp[-S_{ab}(t_e^1)]
 \right.\non
 && \left. +2 r_k \phi_{K}^{P} (x_3, b_3) \alpha_s(t_e^2)
 h_e(1-x_3,x_1,b_3,b_1)\exp[-S_{ab}(t_e^2)] \right\} \;,
 \label{eq:ab}
 \eeq
where  $C_F=4/3$ is a color factor. The function $h_e^i$, the energy scales
$t_e^i$ and the Sudakov factors $S_{ab}$ are displayed in the
Appendix. In the above equation, we do not include the Wilson
coefficients of the corresponding operators, which are process
dependent. They will be shown later in this section for different
decay channels. The diagrams Fig.~1(a) and 1(b) are also the
diagrams for the $B_{s} \to K $ form factor $F_0^{B_{s}\to K}$.
Therefore we can extract $F_0^{B_{s}\to K }$ from Eq.~(\ref{eq:ab}).

\begin{figure}[t,b]
\vspace{-2 cm}
\centerline{\epsfxsize=21 cm \epsffile{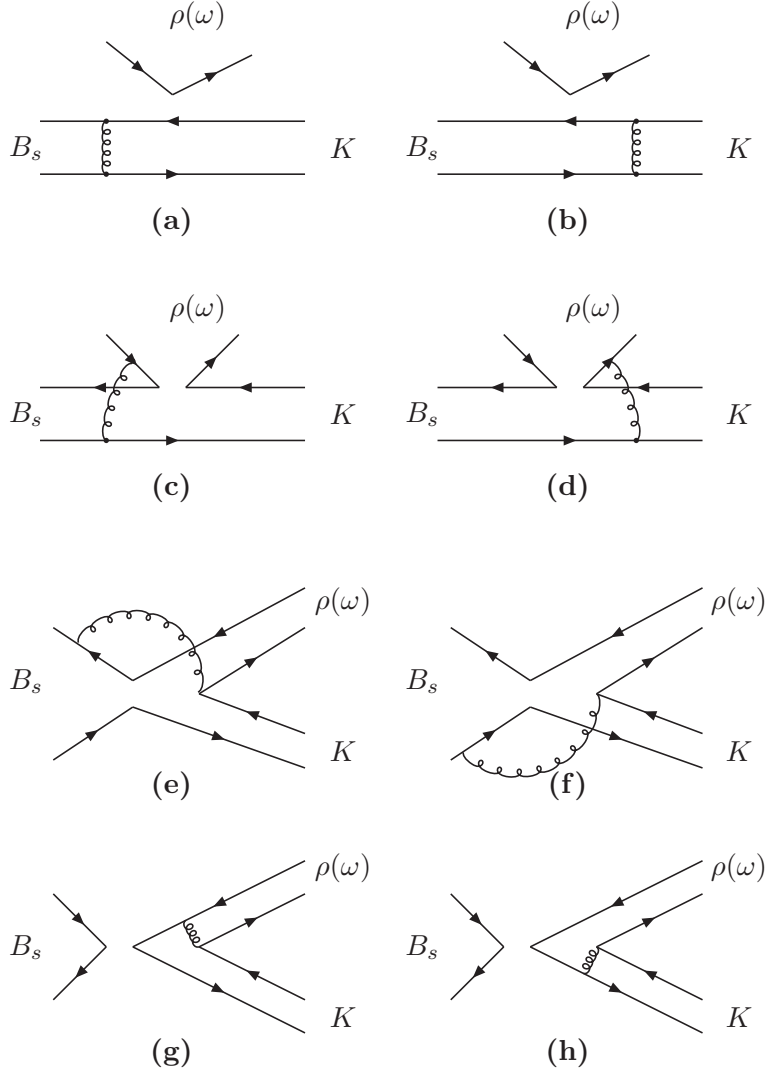}}
\vspace{-14cm}
\caption{ Diagrams contributing to the $B_s\to \rho K$ and $\omega K$
 decays (diagram (a) and (b) contribute to the $B_s\to K$ form
 factor $F_0^{B_s\to K}$). }
 \label{fig:fig1}
\end{figure}

The operators $O_5$, $O_6$, $O_7$, and $O_8$ have a structure of
$(V-A)(V+A)$. In some decay channels, some of these operators
contribute to the decay amplitude in a factorizable way. Since only
the axial-vector part of $(V+A)$ current contribute to the
pseudo-scaler meson production,
 \beq
\langle \rho |V-A|B\rangle \langle K |V+A | 0 \rangle = -\langle
\rho |V-A |B  \rangle \langle K |V-A|0 \rangle  ,
 \eeq
 the result of these $(V-A)(V+A)$ operators is opposite to Eq.~(\ref{eq:ab}), i.e.,
 \beq
 F_{eK}^{P_{1}} &=& -F_{eK} .
\eeq
In some other cases, we need to do Fierz transformation  for
these operators to get right color structure for factorization to
work. In this case, we get $(S+P)(S-P)$ operators from $(V-A)(V+A)$
ones. For these $(S+P)(S-P)$ operators, Fig.~ 1(a) and 1(b)
will give
\beq
F_{eK}^{P_{2}}&=& 0  \; .
\eeq

For the non-factorizable diagram 1(c) and 1(d), all three meson wave
functions are involved. The integration of $b_3$ can be performed
using $\delta$ function $\delta(b_3-b_1)$, leaving only integration
of $b_1$ and $b_2$. For the $(V-A)(V-A)$ operators, the result is:
\beq
M_{eK}& = & \frac{32} {\sqrt{6}} \pi C_F  M_{B_{s}}^{4}
\int_{0}^{1}d x_{1}d x_{2}\,d x_{3}\,\int_{0}^{\infty} b_1d b_1 b_2d
b_2\, \phi_{B_{s}}(x_1,b_1) \phi_{\rho}(x_2,b_2) \non
 & &\cdot
\left. \{ (1-x_3) \left[\phi_{K}^{A}(x_3,b_1)+2 r_k
\phi_{K}^{T}(x_3,b_1)\right]\right.\non
  & &\cdot
\left. \alpha_s(t_f)
 h_f(x_1,x_2,1-x_3,b_1,b_2)\exp[-S_{cd}(t_f)]\right. \} \; .
 \eeq

For the $(V-A)(V+A)$ operators, the result is:
\beq
M_{eK}^{P_{1}}& = & \frac{64} {\sqrt{6}} \pi C_F G_F M_{B_{s}}^{4}r_{\rho}
\int_{0}^{1}d x_{1}d x_{2}\,d x_{3}\,\int_{0}^{\infty} b_1 d b_1 b_2 d
b_2\, \phi_{B_{s}}(x_1,b_1) \non & &\cdot\left
\{\left[x_{2}\phi_{K}^{A}(x_3,b_1)\left(\phi_{\rho}^{s}(x_2,b_2)-
\phi_{\rho}^{t}(x_2,b_2)\right)+r_{K}\left((1+x_{2}-x_{3})
\right.\right.\right. \non
 &&
\left(\phi_{K}^{P}(x_3,b_1)\phi_{\rho}^{s}(x_2,b_2)-
\phi_{K}^{T}(x_3,b_1)\phi_{\rho}^{t}(x_2,b_2)\right)+(1-x_2-x_3)
\non &&
\left.\left.\left(\phi_{K}^{P}(x_3,b_1)\phi_{\rho}^{t}(x_2,b_2)-
\phi_{K}^{T}(x_3,b_1)\phi_{\rho}^{s}(x_2,b_2) \right)\right)\right]
\non &&
 \left.\alpha_s(t_f)
 h_f(x_1,x_2,1-x_3,b_1,b_2)\exp[-S_{cd}(t_f)]\right. \} \; .
 \eeq
For the $(S-P)(S+P)$ operators, the result is:
 \beq
 M_{eK}^{P_{2}} & = & - M_{eK} \; .
 \eeq

For the non-factorizable annihilation diagram 1(e) and 1(f), again
all three wave functions are involved. Here we have two kinds of
contributions. $M_{ak}$ is the contribution containing operator type
$(V-A)(V-A)$, while $M_{aK}^{P_{1}}$ is the contribution containing
operator type $(V-A)(V+A)$:
 \beq
   M_{aK}&=& \frac{32} {\sqrt{6}} \pi C_F M_{B_{s}}^{4}
 \int_{0}^{1}d x_{1}d x_{2}\,d x_{3}\,\int_{0}^{\infty} b_1d b_1 b_2d
   b_2\, \phi_{B_{s}}(x_1,b_1)\non
  && \cdot
   \left\{ \left[x_2\phi_{\rho}(x_2,b_2)\phi_{K}^{A}(x_3,b_2)+
   r_{\rho}r_{K} \left((x_2-x_3) \right.\right.\right.
  \non
  &&\cdot
  \left( \phi_{K}^{P}(x_3,b_2)\phi_{\rho}^{t}(x_2,b_2)+
   \phi_{K}^{T}(x_3,b_2)\phi_{\rho}^{s}(x_2,b_2)\right)+
   (2+x2+x3)
  \non
  && \cdot
  \left.\left.\phi_{K}^{P}(x_3,b_2)\phi_{\rho}^{s}(x_2,b_2)
  +(-2+x_2+x_3)\phi_{K}^{T}(x_3,b_2)\phi_{\rho}^{t}(x_2,b_2)\right)\right]
  \non
  && \cdot
 \left.\alpha_s(t_{f}^{2})
 h_{f}^{2}(x_1,x_2,x_3,b_1,b_2)\exp[-S_{ef}(t_{f}^{2})]\right.
  \non
  &&
  +\left[-x_3\phi_{\rho}(x_2,b_2)\phi_{K}^{A}(x_3,b_2)+
  r_{\rho}r_{K} \left((x_2-x_3)\left(
  \phi_{K}^{P}(x_3,b_2)\phi_{\rho}^{t}(x_2,b_2)+\phi_{K}^{T}(x_3,b_2)
  \right.\right.\right.
  \non
  &&\cdot
  \left.\left.\left.\phi_{\rho}^{s}(x_2,b_2)\right) -
  (x_2 +x_3)\left(\phi_{K}^{P}(x_3,b_2)\phi_{\rho}^{s}(x_2,b_2)
  +\phi_{K}^{T}(x_3,b_2)\phi_{\rho}^{t}(x_2,b_2)\right)\right)\right]
  \non
   &&\cdot
 \left.\alpha_s(t_{f}^{1})
 h_{f}^{1}(x_1,x_2,x_3,b_1,b_2)\exp[-S_{ef}(t_{f}^{1})]\right \} \; ,\\
 M_{aK}^{P_{1}}&=& \frac{32} {\sqrt{6}} \pi C_F  M_{B_{s}}^{4}
 \int_{0}^{1}d x_{1}d x_{2}\,d x_{3}\,\int_{0}^{\infty} b_1d b_1 b_2d
   b_2\, \phi_{B_{s}}(x_1,b_1)\non
  && \cdot
   \left\{ \left[r_{\rho}(x_2-2)\phi_{K}^{A}(x_3,b_2)
   \left(\phi_{\rho}^{t}(x_2,b_2)+\phi_{\rho}^{s}(x_2,b_2)\right)
   \right.\right.\non
  &&
  \left.+r_{K}(2-x_3)\phi_{\rho}(x_2,b_2)\left(
  \phi_{K}^{P}(x_3,b_2)+\phi_{K}^{T}(x_3,b_2)\right)\right]\non
  && \cdot
  \left.\alpha_s(t_{f}^{2})
  h_{f}^{2}(x_1,x_2,x_3,b_1,b_2)\exp[-S_{ef}(t_{f}^{2})]\right.
  \non
  &&
   +\left[r_{\rho}(-x_2)\phi_{K}^{A}(x_3,b_2)
   \left(\phi_{\rho}^{t}(x_2,b_2)+\phi_{\rho}^{s}(x_2,b_2)\right)
   \right.\non
  &&
  \left.+r_{K} x_3 \phi_{\rho}(x_2,b_2)\left(
  \phi_{K}^{P}(x_3,b_2)+\phi_{K}^{T}(x_3,b_2)\right)\right]\non
  &&\cdot
 \left.\alpha_s(t_{f}^{1})
 h_{f}^{1}(x_1,x_2,x_3,b_1,b_2)\exp[-S_{ef}(t_{f}^{1})]\right \} \; ,
 \eeq
where $r_K = m_0^{K}/M_{B_{s}} $ and $r_\rho = m_\rho/M_{B_{s}}$.

The factorizable annihilation diagram 1(g) and 1(h) involve only
$\rho$ and $K$ wave functions. There are also two kinds of decay
amplitudes for these two diagrams. $F_{aK}$ is for $(V-A)(V-A)$ type
operators, and $F_{aK}^{P_{2}}$ is for $(S-P)(S+P)$ type operators,
\beq
  F_{aK} &=& 8 \pi C_F  f_{B_{s}} M_{B_{s}}^{4}\int_{0}^{1}d
 x_{2}\,d x_{3} \,\int_{0}^{\infty} b_2d b_2b_3d b_3 \,
 \non
  & &\cdot
  \left\{ \left[ x_3 \phi_\rho(x_2,b_2)
  \phi_{K}^{A}(x_3,b_3) + 2 r_\rho r_k \phi_{\rho}^{s}(x_2,b_2)
  \left((1+x_3)\phi_{K}^{P}(x_3,b_3)\right.\right.\right. \non
   &&
   \left.\left.\left.(x_3-1)\phi_{K}^{T}(x_3,b_3)\right)\right]
   \alpha_s(t_e^3)h_a(x_2,x_3,b_2,b_3)\exp[-S_{gh}(t_e^3)]\right.
   \non
   &&
   \left.-\left[x_2 \phi_\rho(x_2,b_2)
  \phi_{K}^{A}(x_3,b_3) + 2 r_\rho r_k \phi_{K}^{P}(x_3,b_3)
  \left((1+x_2)\phi_{\rho}^{s}(x_2,b_2)\right.\right.\right. \non
   &&
    \left.\left.\left. (x_2-1)\phi_{\rho}^{t}(x_2,b_2)\right)\right]
   \alpha_s(t_e^4)h_a(x_3,x_2,b_3,b_2)\exp[-S_{gh}(t_e^4)]\right \}\;,
\eeq
 %%%%%%%%%%%%%%%%%%%%%%%%%%%%%%%%%%%
\beq
 F_{aK}^{P_{2}} &=& -16 \pi C_F  f_{B_{s}} M_{B_{s}}^{4}\int_{0}^{1}d
  x_{2}\,d x_{3} \,\int_{0}^{\infty} b_2d b_2b_3d b_3 \,
  \non
  & &\cdot\left\{ \left[2 r_{\rho}\phi_{\rho}^{s}(x_2,b_2)
   \phi_{K}^{A}(x_3,b_3)+r_k x_3 \phi_{\rho}(x_2,b_2)
   \left( \phi_{K}^{P}(x_3,b_3)\right.\right.\right.\non
   &&
    \left.\left.\left.-\phi_{K}^{T}(x_3,b_3)\right)\right]
    \alpha_s(t_e^3)h_a(x_2,x_3,b_2,b_3)\exp[-S_{gh}(t_e^3)]\right.
   \non
   &&
   +\left[x_2 r_\rho \phi_{K}^{A}(x_3,b_3)\left(\phi_{\rho}^{s}(x_2,b_2)
    -\phi_{\rho}^{t}(x_2,b_2)\right)+2 r_k
    \phi_{\rho}(x_2,b_2)\right.\non
   & &\cdot
     \left. \phi_{K}^{P}(x_3,b_3)\alpha_s(t_e^4)
    h_a(x_3,x_2,b_3,b_2)\exp[-S_{gh}(t_e^4)]\right \}\;.
\eeq

In the above equations, we have  assumed that $x_1 <<x_2,x_3$. Since
the light quark momentum fraction $x_1$ in $B$ meson is peaked at
the small $x_1$ region, this is not a bad approximation. The
numerical results also show that this approximation makes very
little difference in the final result. After using this
approximation, all the diagrams are functions of $k_1^-= x_1
M_{B_{s}}/\sqrt{2}$ of $B_{s}$ meson only, independent of the
variable of $k_1^+$. Therefore the integration of eq.(\ref{int}) is
performed safely.

%%%%%%%%%%%%%%%%%%%%%%%%%%%%%%%%%%%%%%%%%%%%%%%%%%%%%%%%%%%%%%%

Combining the contributions from different diagrams, the total decay
amplitude for $B_{s} \to \rho^+ K^{-}$ decay can be written as:
    \beq
{\cal M}(\rho^+ K^{-}) &=&  F_{eK} \left[\xi_u  \left( \frac{1}{3}C_1+C_2\right)
    -\xi_t\left( \frac{1}{3}C_3+C_4+\frac{1}{3}C_9+C_{10}\right)\right]  \non
   & &
  + M_{eK}\left[\xi_u\left(C_1\right)
  -\xi_t\left(C_3+C_9\right)\right] + M_{eK}^{P_{1}}\left[-\xi_t\left(C_5+C_7\right)\right]  \non
   & &
 +M_{aK}\left[-\xi_t\left(C_3-\frac{1}{2}C_9\right)\right]
   +M_{aK}^{P_{1}}\left[-\xi_t\left(C_5-\frac{1}{2}C_7\right)\right]  \non
   & &
 +F_{aK}\left[-\xi_t\left(\frac{1}{3}C_3+C_4-\frac{1}{6}C_9-\frac{1}{2}C_{10}
   \right)\right]  \non
   & &
 +F_{aK}^{P_{2}}\left[-\xi_t\left(\frac{1}{3}C_5+C_6-\frac{1}{6}C_7-\frac{1}{2}C_8
   \right)\right   ],
   \label{eq:m1}
\eeq
where $\xi_u = V_{ub}^*V_{ud}$, $\xi_t = V_{tb}^*V_{td}$¡£

Similarly, the decay amplitude for $B_{s} \to \rho^0 \overline{K}^0$ can be written as:
\beq
{\cal M}(\rho^0 \overline{K}^0) &=& F_{eK} \left[\xi_u\left(C_1+\frac{1}{3}C_2\right)
    -\xi_t\left(-\frac{1}{3}C_3-C_4+\frac{3}{2}C_7+\frac{1}{2}C_8
    +\frac{5}{3}C_9+C_{10}\right)\right] f_1  \non
& &
+ M_{eK}\left[\xi_u C_2
    -\xi_t\left( -C_3-\frac{3}{2}C_8+\frac{1}{2}C_9+\frac{3}{2}C_{10}
    \right)\right] f_1 \non
    & &
+ M_{eK}^{P_{1}}\left[-\xi_t\left(-C_5+\frac{1}{2}C_7\right)    \right] f_1
+ M_{aK}\left[-\xi_t\left(-C_3+\frac{1}{2}C_9\right)\right]f_1  \non
   & &
+ M_{aK}^{P_{1}}\left[-\xi_t\left(-C_5+\frac{1}{2}C_7\right)\right] f_1
+ F_{aK}\left[-\xi_t\left(-\frac{1}{3}C_3-C_4+\frac{1}{6}
    C_9+\frac{1}{2}C_{10}\right)\right] f_1  \non
    & &
+ F_{aK}^{P_{2}}\left[-\xi_t\left(-\frac{1}{3}C_5-C_6
     +\frac{1}{6}C_7+\frac{1}{2}C_8\right) \right  ] f_1 ,
     \label{eq:m2}
\eeq
where $f_1=1/\sqrt{2}$.

%%%%%%%%%%%%%%%%%%%%%%%%%%%%%%%%%%%%%%%%%
For the decay amplitude of $B_{s} \to \omega   \overline{K}^0$ decay,
one can obtain its decay amplitude from Eq.~(\ref{eq:m2}) by replacing  the vector meson $\rho$
with $\omega$, i.e,
\beq
    f_\rho \to f_\omega,  \quad    f_\rho^T \to f_\omega^T, \quad    m_\rho \to  m_\omega.
\eeq
Note that we have considered  the difference in the quark components
for the two scalar mesons $K^+$ and $K^0$ in the analytic expressions.
We denote the corresponding amplitudes for $B_{s} \to \omega   \overline{K}^0$ decay
as $F_{eK}^\prime $, $M_{eK}^\prime $,
$M_{eK}^{\prime P_{1}}$, $M_{aK}^\prime $, $M_{aK}^{\prime P_{1}}$,
 $F_{aK}^\prime $ and $F_{aK}^{\prime P_{2}}$,  but do not show explicit expressions
of these amplitudes here for the sake of simplicity.
The total amplitude finally can be written as
   \beq
{\cal M}(\omega \overline{K}^0) &=& F_{eK}^\prime
    \left[\xi_u\left(C_1+\frac{1}{3}C_2\right)f_{1}
    -\xi_t\left(\frac{7}{3}C_3+\frac{5}{3}C_4 \right.\right.\non
    & &
    \left.\left.+2C_5+\frac{2}{3}C_6+\frac{1}{2}C_7+\frac{1}{6}C_8
    +\frac{1}{3}C_9-\frac{1}{3}C_{10}\right)f_{1}\right]\non
    & &
    +M_{eK}^{'}\left[\xi_u\left(C_2\right)f_{1}
    - \xi_t\left(C_3+2C_4-2C_6-\frac{1}{2}C_8-\frac{1}{2}C_9
    +\frac{1}{2}C_{10}\right)f_{1}\right]\non
    & &
    - M_{eK}^{'P_{1}}\; \xi_t\left(C_5-\frac{1}{2}C_7
    \right)f_{1} - M_{aK}^{'} \; \xi_t\left(C_3-\frac{1}{2}C_9
    \right)f_{1} \non
    & &
    - M_{aK}^{'P_{1}} \; \xi_t\left(C_5-\frac{1}{2}C_7
    \right)f_{1}  - F_{aK}^{'} \; \xi_t\left(\frac{1}{3}C_3
    +C_4-\frac{1}{6}C_9-\frac{1}{2}C_{10} \right)f_{1} \non
    & &
    +F_{aK}^{'P_{2}} \; \xi_t\left(\frac{1}{3}C_5
    +C_6-\frac{1}{6}C_7-\frac{1}{2}C_8  \right)f_{1} , \label{eq:m3}
    \eeq
where $f_1=1/\sqrt{2}$\;.

\section{Numerical results and Discussions}\label{sec:n-d}

\subsection{Input parameters and wave functions}

In the numerical calculations we use the following input parameters
 \beq
  \Lambda_{\overline{\mathrm{MS}}}^{(f=4)} &=& 250 {\rm MeV}, \quad
  f_\rho = 205 {\rm MeV},\quad f_\rho^T = 160 {\rm MeV}, \non
  m_0^K &=& 1.6 {\rm GeV}, \quad f_{B_{s}} = 236 {\rm MeV},
  \quad f_K = 160  {\rm MeV}, \non
  m_\omega &=& 0.782 {\rm GeV},\quad f_\omega = 200 {\rm MeV},
  \quad f_\omega^T = 160 {\rm MeV}, \non
  m_\rho &=& 0.770 {\rm GeV},\quad M_{B_{s}} = 5.37 {\rm GeV},
  \quad M_W = 80.42 {\rm GeV}.
  \label{para}
  \eeq
  The central values of the CKM matrix elements to be used in
  numerical calculations are \cite{pdg04}
  \beq
  |V_{ud}|= 0.9745,    \quad |V_{ub}|=0.0038, \quad |V_{tb}|= 1, \quad |V_{td}|=0.0083.
  \eeq

 For the $B_{s}$ meson wave function, we adopt the model
  \beq
 \phi_{B_{s}}(x,b) &=&  N_{B_{s}} x^2(1-x)^2 \mathrm{exp} \left
 [ -\frac{M_{B_{s}}^{2}\ x^2}{2 \omega_{b}^2} -\frac{1}{2} (\omega_{b} b)^2\right],
 \label{phib}
 \eeq
 where $\omega_{b}$ is a free parameter and we take
 $\omega_{b}=0.5\pm 0.05$ GeV in numerical calculations, and
 $N_{B_s}=65.332$ is the normalization factor for $\omega_{b}=0.5$.
 This is the same wave function as used in  Refs.~\cite{sunbs03,pipi,xql15}.

For the light meson wave function, we neglect the $b$ dependant
part, which is not important in numerical analysis. We choose the
wave function of $\rho$ and $\omega$ meson as given in Ref.~\cite{pb23}
  \beq
  \phi_{\rho(\omega)}(x) &=& \frac{3}{\sqrt{6} }
  f_{\rho(\omega)}  x (1-x)  \left[1+ 0.18C_2^{3/2} (2x-1) \right],\\
    \phi_{\rho(\omega)}^t(x) &=&  \frac{f_{\rho(\omega)}^T }{2\sqrt{6} }
  \left\{  3 (2 x-1)^2 +0.3(2 x-1)^2  \left[5(2 x-1)^2-3  \right]
  \right.
  \nonumber\\
  &&~~\left. +0.21 [3- 30 (2 x-1)^2 +35 (2 x-1)^4] \right\},\\
  \phi_{\rho(\omega)}^s(x) &=&  \frac{3}{2\sqrt{6} }
  f_{\rho(\omega)}^T   (1-2x)  \left[1+ 0.76 (10 x^2 -10 x +1) \right] .
 \eeq
 The Gegenbauer polynomial is defined by
 \beq
 C_2^{3/2} (t) = \frac{3}{2} \left (5t^2-1 \right ).
 \eeq

For $K$ meson's wave functions, $\phi_K^A$, $\phi_K^P$ and $\phi_K^T$ describe
the axial vector, pseudoscalar and tensor components respectively. We utilize the result
from the light-cone sum rule~\cite{ak02} including twist-3 contribution:
 \beq
 \phi_K^A(x)&=&\frac{3}{\sqrt{2N_c}}f_K x(1-x)
 \left[1+0.15t+0.405\left(5t^2-1\right)\right]\;,\non
\phi^P_K(x)&=&\frac{1}{2\sqrt{2N_c}}f_K
 \left[1+0.106\left(3t^2-1\right)-\frac{0.148}{8}\left(3-30t^2+35t^4\right)\right]
 \;,\non
\phi^T_K(x)&=&\frac{1}{2\sqrt{2N_c}}f_K \; t   \left[1+0.1581\left(5t^2-3\right)\right]\;.
\eeq
 whose coefficients correspond to $m_0^K=1.6$ GeV and $t=1-2x$. The  parameter $m_0^K$ is defined as:
 \beq
 m_0^K \equiv
 \frac{m_K^2}{(m_u+m_s)}\;.
 \label{eq:19}
 \eeq
For the pseudoscalar $K$ meson we  have assumed that the wave function
of $\overline{K^0}$ is the same as the wave function of
$K^-$, since one can assume that the mass of the u quark is equivalent to
the mass of the d quark due to the isospin symmetry.

We include full expression of twist$-3$ wave functions for light mesons.
The twist$-3$ wave functions are also adopted from QCD sum rule  calculations~\cite{vmb39}.
Using the above chosen wave functions and the relevant input parameters,
we find the numerical value of
the corresponding form factor at zero momentum transfer from Eqs.(\ref{eq:ab}):
 \beq
 F_0^{B_s \to K}(q^2=0)&=& \frac{F_{eK}}{m_{B_s}^2}=0.276 ^{+0.050}_{-0.040} \;,
 \label{eq:aff0}
\eeq
for $\omega_b=0.50 \pm 0.05$ GeV, which agrees well with the value as given in
Refs.~\cite{yu-7107,dsd00}.

\subsection{Branching ratios}

For $B_{s} \to \rho(\omega) K$ decays, the decay amplitudes as given in
Eqs.~(\ref{eq:m1}), (\ref{eq:m2}) and (\ref{eq:m3}) can be
rewritten as
 \beq
 {\cal M} &=& V_{ub}^*V_{ud} T -V_{tb}^* V_{td} P= V_{ub}^*V_{ud} T
 \left [ 1 + z e^{ i ( \alpha + \delta ) } \right],
 \label{eq:ma}
 \eeq
 where $\alpha = \arg \left[-\frac{V_{td}V_{tb}^*}{V_{ud}V_{ub}^*}\right]$ is the weak
phase (one of the three CKM angles), and $\delta$ is the relative strong phase
between the tree ( ``T" ) and  penguin (``P" ) amplitude, while the term ``z" describes
the ratio of penguin to tree contributions and is defined as
\beq
 z=\left|\frac{V_{tb}^* V_{td}}{ V_{ub}^*V_{ud} } \right|
 \left|\frac{P}{T}\right|.
\label{eq:zz}
\eeq
The ratio $z$ and the strong phase $\delta$ can be calculated in the pQCD
approach. For  $B_{s} \to \rho^+ K^{-}$ , $\rho^0 \overline{K}^0$  and $\omega
\overline{K}^0$ decays, we find numerically that
\beq
 z(\rho^+ K^{-}) &=& 0.10, \quad \delta (\rho^+ K^{-}) = + 138^\circ , \label{eq:zd1}\\
 z(\rho^0\overline{K}^0) &=& 1.5 , \quad  \delta(\rho^0\overline{K}^0) = +77^\circ ,\label{eq:zd2}\\
 z(\omega\overline{K}^0) &=& 1.9 ,\quad  \delta(\omega\overline{K}^0)= +273^\circ .\label{eq:zd3}
\eeq
For $B_{s} \to \rho\overline{K}^0$ and $B_{s}\to
\omega\overline{K}^0$ decays, the "Tree" diagram  contribution is
suppressed by a factor of $C_1 + C_2/3 \sim 0.1$.
Thus the penguin diagram contribution is comparable with  the tree
contribution. In our
calculation, the only input parameters are wave functions,which stand
for the non-perturbative contributions. Up to now, no exact solution
is made for them. So the main uncertainty in pQCD approach comes from
these wave functions. In this paper, we choose
the light cone wave functions which are obtained from QCD Sum Rules
\cite{ak02,vmb39}. For heavy B and $B_{s}$ mesons, its wave function
is still under discussion using different approaches. In this paper,
we find the branching ratio of $B_{s} \to \rho(\omega) K$ is
sensitive to the wave function parameter $\omega_{b}$. So the main
errors of the ratio $z$ and the strong phase $\delta$ are induced by
the uncertainty of $\omega_b =0.5 \pm 0.05$ GeV. We just use the
central values of $z$ and $\delta$ in the following numerical calculations, unless stated
explicitly.

From Eq.~(\ref{eq:ma}), it is easy to write the decay amplitude $ \overline{\cal M} $
for the corresponding charge conjugated decay mode. And the CP-averaged branching ratio
for considered decays is generally defined as
\beq
 Br = (|{\cal M}|^2 +|\overline{\cal M}|^2)/2 =  \left|
 V_{ub}V_{ud}^* T \right| ^2 \left[1 +2 z\cos \alpha \cos \delta
 +z^2 \right], \label{br}
\eeq
where the ratio $z$ and the strong phase $\delta$ have been defined
in Eqs.(\ref{eq:ma}) and (\ref{eq:zz}).

Using  the wave functions and the input parameters as specified in
previous sections,  it is straightforward  to calculate the CP
averaged branching ratios for the three considered decays. The
theoretical predictions in the pQCD approach for the branching
ratios of the decays under consideration are the following:
\beq
 Br(B_{s}\to \rho^{\pm} K^{\mp}) &=& \left [24.7^{+10.1}_{-6.7}(\omega_b)
    ^{+1.1}_{-1.2}  (\alpha )\right ]  \times 10^{-6},\\
  Br(B_{s} \to \rho^{0} \overline{K}^0)  &=& \left [1.2^{+0.4}_{-0.2}
   (\omega_b) \pm 0.1 (\alpha )\right ]  \times 10^{-7},\\
  Br(B_{s} \to \omega \overline{K}^0)  &=& \left [1.7^{+0.6}_{-0.3}
   (\omega_b)\pm 0.02 (\alpha ) \right ] \times 10^{-7},
\eeq
where the major errors are induced by the uncertainty of $\omega_b=0.5 \pm 0.05$ GeV, and
$\alpha =100^\circ \pm 20^\circ$, respectively. It is easy to see that the above pQCD
predictions for the branching ratios are  consistent with those
predicted in the QCD factorization approach \cite{mb333},
\beq
   Br(B_{s} \to \rho^{\pm} K^{\mp}) &=& \left [24.5^{+15.2}_{-12.9}
   \right ]  \times 10^{-6} , \\
   Br(B_{s} \to \rho^{0} \overline{K}^0)  &=& \left [6.1^{+12.6}_{-6.0}
   \right ]  \times 10^{-7} , \\
   Br(B_{s} \to \omega \overline{K}^0)  &=& \left [5.1^{+8.3}_{-4.0}
    \right ]  \times 10^{-7} ,
    \eeq
within one standard deviation, although the central values of pQCD predictions for
$B_s \to (\rho^0, \omega) \overline{K}^0$ decays are much smaller than those in QCD factorization approach.

 \begin{figure}[tb]
 \centerline{\mbox{\epsfxsize=10cm\epsffile{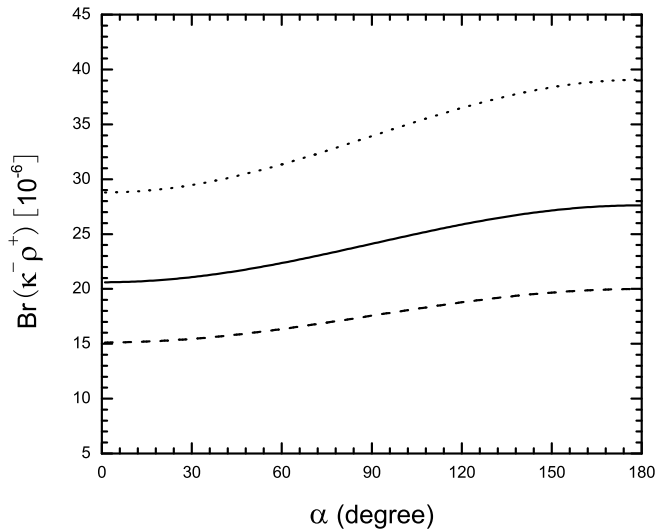}}}
 \vspace{0.3cm}
 \caption{The branching ratio (in unit of $10^{-6}$)
 of $B_{s}\to \rho^{\pm} K^{\mp}$  decay for  $\omega_b=0.45 $ GeV
 (dotted curve), $0.50$ GeV (solid  curve) and $0.55$ GeV (dashed curve),
  as a function of the CKM angle $\alpha$.}
 \label{fig:fig2}
 \end{figure}

\begin{figure}[tb]
\centerline{\mbox{\epsfxsize=9cm\epsffile{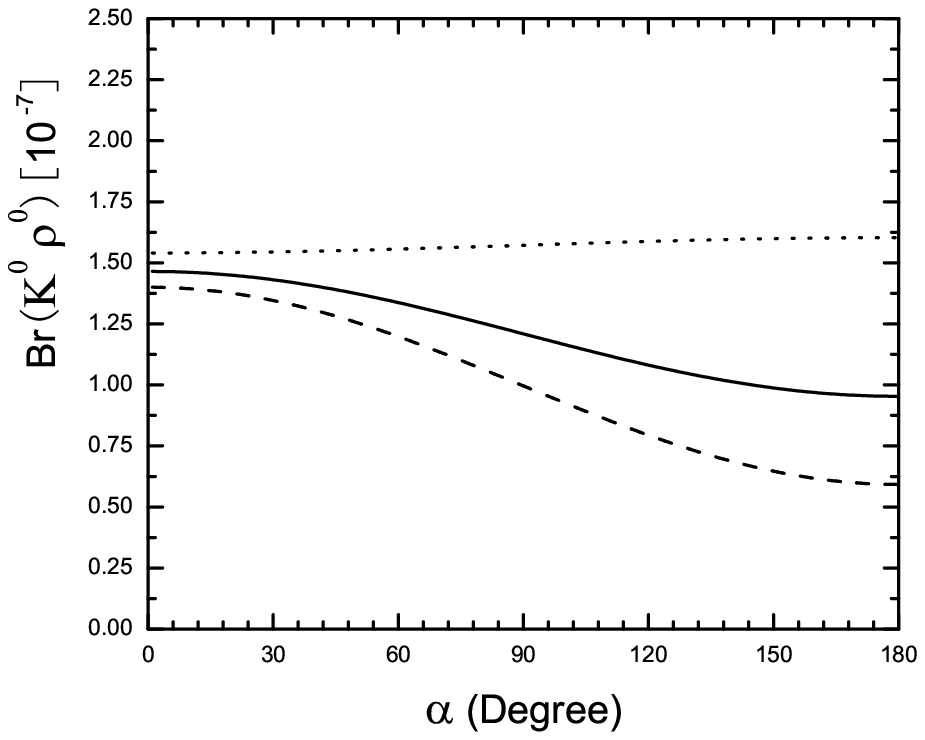}\epsfxsize=9cm\epsffile{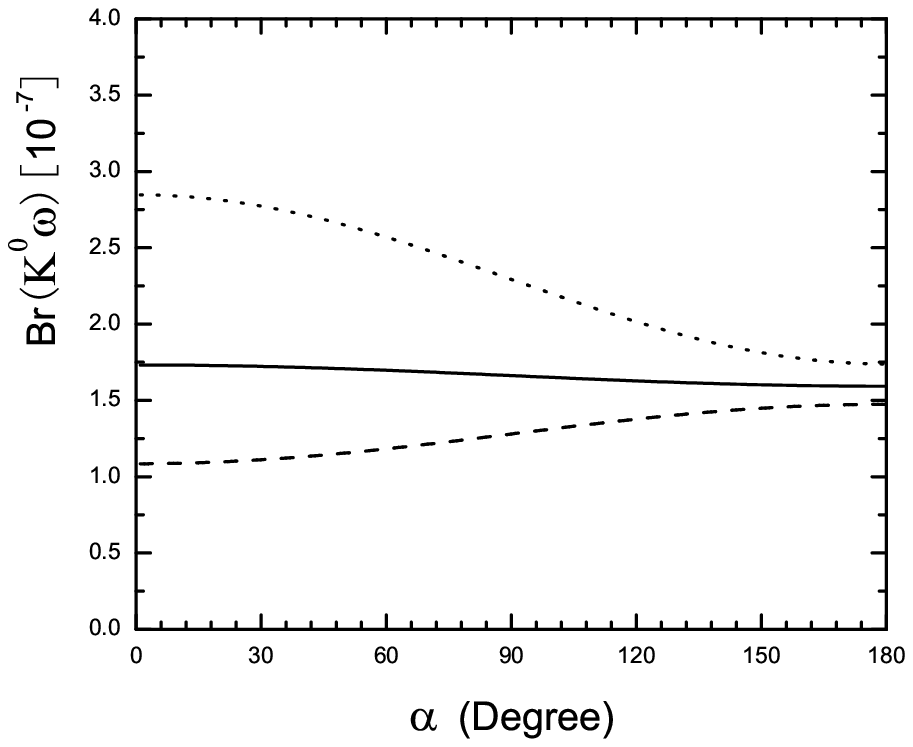}}}
\vspace{0.3cm}
\caption{The branching ratio (in unit of $10^{-7}$) of $B_{s} \to \rho^0 \overline{K}^0 $ and
$\omega \overline{K}^0$  decays for  $\omega_b=0.45 $ GeV
 (dotted curve), $0.50$ GeV(solid  curve) and $0.55$ GeV(dashed curve), as a function of the
 CKM angle $\alpha$.}
 \label{fig:fig3}
\end{figure}

In Figs.~\ref{fig:fig2} and \ref{fig:fig3}, we show the $\alpha$-dependence of the
pQCD predictions for the CP averaged  branching ratios of $B_{s} \to \rho^{\pm} K^{\mp}$,
$\rho^0 \overline{K}^0$ and $\omega \overline{K}^0$ decays for
$\alpha= [0^\circ,180^\circ]$ and $\omega_b=0.5\pm 0.05$ GeV.

It is worth stressing  that the theoretical predictions in the pQCD
approach have large theoretical errors induced by the large uncertainties of
parameter $\omega_{b}$ and CKM angle $\alpha$. From Figs.~\ref{fig:fig2} and \ref{fig:fig3},
we observe that the pQCD predictions are sensitive to the variations of $\omega_b$.
This sensitive dependence could be fixed by the well measured $B_{s} \to K$ form
factors from the semi-leptonic $B_{s}$ decays as expected in LHCb experiment.
Other uncertainties in our calculation include the next-to-leading order $\alpha_{s}$ QCD
corrections and higher twist contributions, which need more complicated calculations.
The parameter $m_0^K \approx 1.6$ GeV  characterizes the relative size of twist 3
contribution to twist 2 contribution. Because of the chiral
enhancement of $m_0^K$, the twist 3 contribution become comparable
in size with the twist 2 contribution.

\subsection{CP-violating asymmetries }

Now we turn to the evaluations of the CP-violating asymmetries of
$B_{s} \to \rho(\omega) K$ decays in pQCD approach. For $B_{s} \to \rho^+ K^{-}$ decay,
the direct CP-violating asymmetry $\acp$ can be defined as:
 \beq
{\cal A}_{CP}^{dir} =  \frac{|{\cal M}|^2 - |\overline{\cal M}|^2 }{
 |{\cal M}|^2+|\overline{\cal M}|^2}=
\frac{-2 z \sin \alpha \sin\delta}{1+2 z\cos \alpha \cos \delta
+z^2}, \label{eq:acp1}
 \eeq
where the ratio $z$ and the strong phase $\delta$ have been defined
in previous subsection and are calculable in pQCD approach.

\begin{figure}[tb]
\vspace{-1cm}
\centerline{\epsfxsize=10cm \epsffile{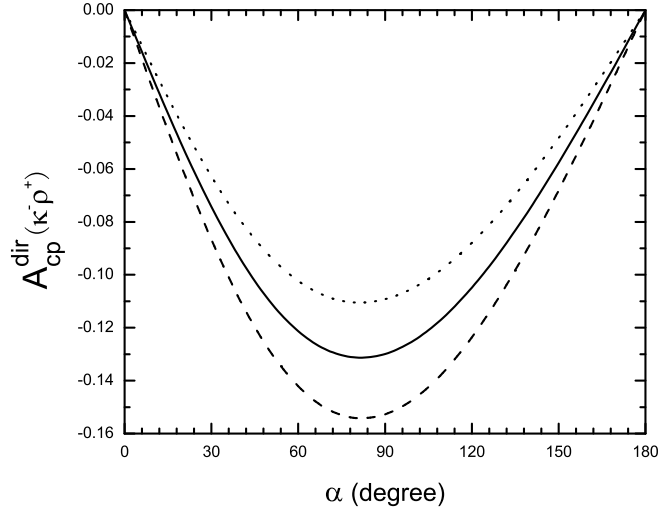}}
\vspace{-0.5cm}
\caption{The $\alpha$-dependence of the direct CP asymmetries  of $B_{s} \to \rho^\pm K^\mp $ decay
for $\omega_b=0.45$ GeV (dotted curve),  $0.50$ GeV (solid curve) and  $0.55$ GeV (dashed curve). }
\label{fig:fig4}
\end{figure}

Using the central values of $z$ and $\delta$ as given in
Eqs.(\ref{eq:zd1}) , (\ref{eq:zd2}) and (\ref{eq:zd3}),  it is easy to
calculate the CP-violating asymmetries:
\beq
\acp^{dir} (B_{s}\to \rho^\pm K^\mp) &=& (-12.5^{+2.0}_{-2.2} (\omega_{b})
^{-0.6}_{+2.0}(\alpha))\times 10^{-2}.
\label{eq:acp1b}
\eeq
Here two major errors are induced by $\omega_b=0.50 \pm 0.05$ GeV and
$\alpha=100^\circ \pm 20^\circ$. The pQCD prediction in Eq.~(\ref{eq:acp1b}) is also
consistent with the prediction in QCDF approach \cite{mb333} within one standard deviation:
$\acp^{dir} (B_{s} \to \rho^\pm K^\mp) = (-1.5 \pm 12.2) \times 10^{-2}$.

In Fig.~\ref{fig:fig4}, we show the $\alpha-$dependence of the direct CP-violating asymmetries
${\cal A}_{CP}^{dir}$ for $B_{s} \to \rho^\pm K^{\mp}$ decay.
The possible theoretical errors induced by the uncertainties of
other input parameters are usually not large, since both $z$ and
$\delta$ are stable against the variations of them. Uncertainties
not included here are the next-to-leading order contributions, which
may affect the CP asymmetry strongly \cite{hnl05}.
For $B_{s} \to \rho^\pm K^{\mp}$ decay, a large CP asymmetry at $10\%$ level plus large branching ratios
at $10^{-5}$ level are clearly measurable in the forthcoming LHCb experiments.

\begin{figure}[tb]
\centerline{\mbox{\epsfxsize=9cm\epsffile{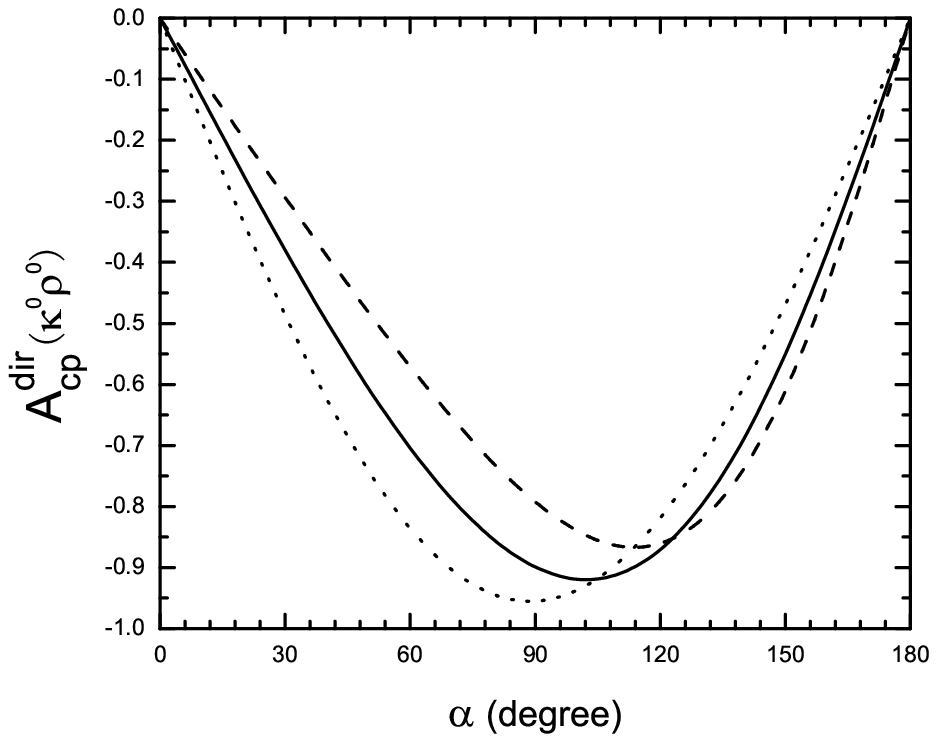}\epsfxsize=9cm\epsffile{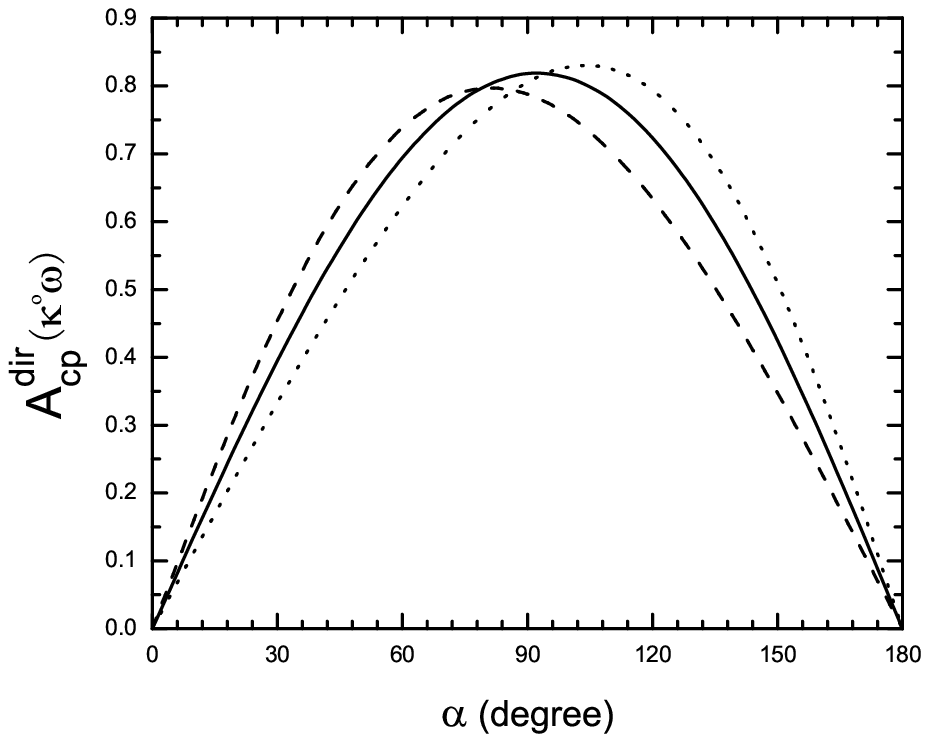}}}
\vspace{0.5cm}
\caption{ The $\alpha$-dependence of the direct CP asymmetries  of $B_{s} \to \rho^0  \overline{K}^0$
and  $\omega\overline{K}^0)$ decays for $\omega_b=0.45$ GeV (dotted curve),  $0.50$ GeV (solid curve)
and  $0.55$ GeV (dashed curve).}
\label{fig:fig5}
\end{figure}

For the pure neutral decays $B_{s} \to \rho^0\overline{K}^0$ and $\omega \overline{K}^0$, there are both
direct and mixing-induced CP violation. Using Eq.~(\ref{eq:acp1}), we find that
\beq
\acp^{dir} (B_{s}\to \rho^0 \overline{K}^0) &=& (-91.9^{-1.8}_{+8.0}(\omega_{b})^{+6.5}_{+4.8}(\alpha))
 \times 10^{-2} ,\non
\acp^{dir} (B_{s} \to \omega \overline{K}^0) &=& (+81.2^{+1.7}_{-5.6}(\omega_{b})^{-1.2}_{-8.8}(\alpha))
\times 10^{-2}, \label{eq:acpd2}
\eeq
for $\omega_b=0.50\pm 0.05$ GeV and $\alpha=100^\circ \pm 20^\circ$.
In Fig.~\ref{fig:fig5}, we show the $\alpha-$dependence of the direct CP-violating asymmetry
for $B_{s} \to \rho^0 \overline{K}^0$ and $B_{s}\to \omega\overline{K}^0$ decays.

The pQCD predictions as given in Eq.~(\ref{eq:acpd2}) are quite different from
those obtained by using the QCDF approach as given in Refs.\cite{mb333} :
\beq
\acp^{dir} (B_{s} \to \rho^0 \overline{K}^0) &=& (24.7^{+58.3}_{-56.8})
  \times 10^{-2} ,\non
\acp^{dir} (B_{s}\to \omega \overline{K}^0) &=& (-43.9^{+69.1}_{-62.1})
  \times 10^{-2}.
\eeq
The reason is the great difference in the source of strong phases in two facterization approaches.
In QCDF approach, the strong phase mainly comes
from the perturbative charm quark loop diagram, which is $\alpha_{s}$
suppressed \cite{mb333}. But the strong phase in pQCD approach comes
mainly from non-factorizable and annihilation type diagrams (see figures 1(c) $ \sim $1 (h) ).
The sign of the direct CP asymmetry is also different for these two approaches for the
latter two neutral decays. The forthcoming LHC-b experiments can make a test for
these two methods.

\begin{figure}[tb]
 \vspace{-1cm} \centerline{\epsfxsize=10cm \epsffile{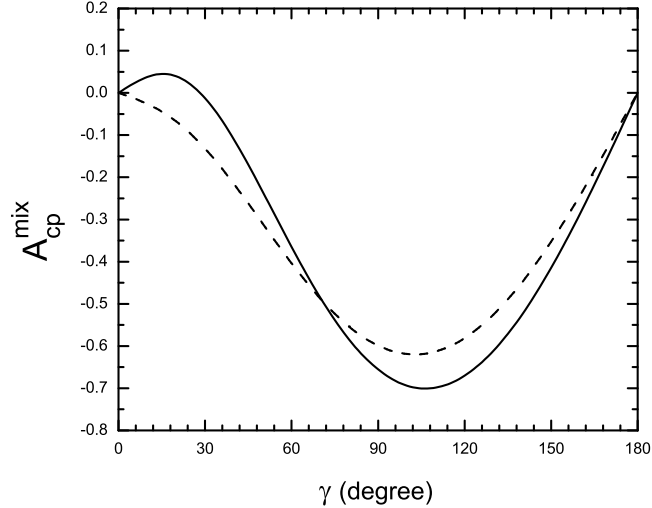}}
 \vspace{-0.5cm}
 \caption{The $\gamma$-dependence of the mixing-induced CP violating asymmetry of  $B_{s}\to \rho^0
 \overline{K}^0 $   (solid curve) and $B_{s} \to \omega\overline{K}^0$ (dashed curve)
                    for fixed $\omega_b=0.50$ GeV.}
 \label{fig:fig6}
 \end{figure}

Following Ref.\cite{yu-7107}, the mixing-induced CP asymmetry for
$B_{s} \to \rho^0 \overline{K}^0$ and $\omega\overline{K}^0$ decays,
can be defined as
\beq
\acp^{mix} =\frac{-2Im(\lambda_{CP})}{1+|\lambda_{CP}|^{2}} =
\frac{\sin 2\gamma +2 Re(x) \sin \gamma }{1+|x|^2 + 2 Re(x)\cos \gamma },
\label{cpf}
\eeq
 where $x= \frac{V_{cb} V_{cd}^*}{|V_{ub}V_{ud}^*|}\frac{P}{T+P}$, and the angle $\gamma$ is one
 of the three CKM angles.  Numerically, the pQCD predictions for the mixing induced CP asymmetry are
\beq
\acp^{mix} (B_{s} \to \rho^0\overline{K}^0) &=& (-37^{+22}_{-19}(\omega_{b})
^{+26}_{+22}(\gamma)) \times 10^{-2} ,\non
\acp^{mix} (B_{s} \to \omega \overline{K}^0) &=&
(-40 \pm 11  (\omega_{b})^{+19}_{-15}(\gamma)) \times 10^{-2},
\eeq
for $\omega_b=0.50\pm 0.05$ GeV and $\gamma=60^\circ \pm 20^\circ$.
The $\gamma$-dependence of the mixing-induced CP asymmetries are shown in Fig.~\ref{fig:fig6}

\section{summary }

In this paper,  we calculate the branching ratios and CP-violating
asymmetries of $B_{s} \to \rho^{\pm} K^{\mp}$, $\rho^{0} \overline{K}^0$ and
$\omega \overline{K}^0$ decays in the pQCD factorization approach.

Besides the usual factorizable diagrams, the non-factorizable and
annihilation diagrams as shown in Fig.~\ref{fig:fig1} are also
calculated analytically. The non-factorizable and
annihilation contributions provide the necessary strong phase required by a
non-zero CP-violating asymmetry for the considered decays.

From our calculations and phenomenological analysis, we found the following
results:
\begin{itemize}

\item
From analytical calculations, the form factor $F^{B_{s}\to K} (0)$ can be extracted.
The pQCD prediction is $F^{B_{s} \to K}(0)=
0.276^{+0.050}_{-0.040}$ for $\omega_{b}=0.50\pm 0.05$ GeV, which agrees well with
the value as given in Ref.~\cite{yu-7107} and the result obtained from the QCD sum rule
calculations \cite{dsd00}.

\item
For the CP-averaged branching ratios of the three considered decay modes, the
pQCD predictions are
 \beq
Br(B_{s} \to \rho^{\pm} K^{\mp}) & \approx &24.7 \times 10^{-6}, \non
Br(B_{s} \to \rho^0 \overline{K}^0) & \approx & 1.2 \times 10^{-7},\non
Br(B_{s} \to \omega \overline{K}^0)  & \approx & 1.7 \times10^{-7}.
\eeq
The theoretical uncertainties are around thirty to fifty percent. The leading pQCD
predictions for the branching ratios  are  also consistent with those obtained by
employing the QCD factorization approach.

\item
For the CP-violating asymmetries, the theoretical predictions in pQCD approach are
\beq
\acp^{dir}(B_{s} \to \rho^\pm K^\mp ) &\approx & -12\% , \non
\acp^{dir}(B_{s} \to \rho^0 \overline{K}^0 ) & \approx & -92\%, \quad
\acp^{mix}(B_{s} \to \rho^0 \overline{K}^0 )  \approx  -36\%, \non
\acp^{dir}(B_{s} \to \omega \overline{K}^0 ) & \approx & +81\%, \quad
\acp^{mix}(B_{s} \to \omega \overline{K}^0 ) \approx -40\%,
\eeq
for $\alpha \approx 100^\circ$, $\gamma \approx 60^\circ$. Of course, the theoretical errors here are
still large.

\end{itemize}

\begin{acknowledgments}

We are very grateful to Cai-Dian L\"u, Libo Guo, Hui-sheng Wang, Xin Liu and Qian-gui Xu
for helpful discussions. This work is partly supported  by the National Natural Science
Foundation of China under Grant No.10275035 and 10575052, and by the
Specialized Research Fund for the Doctoral Program of Higher Education (SRFDP)
under Grant No.~20050319008.

\end{acknowledgments}

%%%%%%%%%%%%%%%%%%%%%%%%%%%%%%%%%%%%%%%%%%%%%%%%%%%%%%%%%%%%%%%%%%%%%%%%%%%%%%%%%%
%                                        Appendix
%%%%%%%%%%%%%%%%%%%%%%%%%%%%%%%%%%%%%%%%%%%%%%%%%%%%%%%%%%%%%%%%%%%%%%%%%%%%%%%%5

\begin{appendix}

\section{Related Functions }\label{sec:aa}

We show here the function $h_i$'s appeared in the expressions of the decay amplitudes in Sec.~\ref{sec:p-c},
coming from the Fourier transformations  of function $H^{(0)}$,
 \beq
 h_e(x_1,x_3,b_1,b_3)&=&
 K_{0}\left(\sqrt{x_1 x_3} m_{B_{s}} b_1\right)
 \left[\theta(b_1-b_3)K_0\left(\sqrt{x_3} m_{B_{s}}
b_1\right)I_0\left(\sqrt{x_3} m_{B_{s}}b_3\right)\right.
 \non
& &\;\left. +\theta(b_3-b_1)K_0\left(\sqrt{x_3}  m_{B_{s}}
b_3\right) I_0\left(\sqrt{x_3}  m_{B_{s}} b_1\right)\right]
S_t(x_3), \label{he1}
 \eeq
 \beq
 h_a(x_2,x_3,b_2,b_3)&=&
 K_{0}\left(i \sqrt{x_2 x_3} m_{B_{s}} b_2\right)
 \left[\theta(b_3-b_2)K_0\left(i \sqrt{x_3} m_{B_{s}}
b_3\right)I_0\left(i \sqrt{x_3} m_{B_{s}} b_2\right)\right.
 \non
& &\;\;\;\;\left. +\theta(b_2-b_3)K_0\left(i \sqrt{x_3}  m_{B_{s}}
b_2\right) I_0\left(i \sqrt{x_3}  m_{B_{s}} b_3\right)\right]
S_t(x_3), \label{he3} \eeq
 \beq
 h_{f}(x_1,x_2,x_3,b_1,b_2) &=&
 \biggl\{\theta(b_2-b_1) \mathrm{I}_0(M_{B_{s}}\sqrt{x_1 x_3} b_1)
 \mathrm{K}_0(M_{B_{s}}\sqrt{x_1 x_3} b_2)
 \non
&+ & (b_1 \leftrightarrow b_2) \biggr\}  \cdot\left(
\begin{matrix}
 \mathrm{K}_0(M_{B_{s}} F_{(1)} b_2), & \text{for}\quad F^2_{(1)}>0 \\
 \frac{\pi i}{2} \mathrm{H}_0^{(1)}(M_{B_{s}}\sqrt{|F^2_{(1)}|}\ b_2) ,&
 \text{for}\quad F^2_{(1)}<0
\end{matrix}
\right),
\label{eq:pp1}
 \eeq
 \beq
 h_f^1(x_1,x_2,x_3,b_1,b_2) &=&
 \biggl\{\theta(b_1-b_2) \mathrm{K}_0(i \sqrt{x_2 x_3} b_1 M_{B_{s}})
 \mathrm{I}_0(i \sqrt{x_2 x_3} b_2 M_{B_{s}})
 \non
&+& (b_1 \leftrightarrow b_2) \biggr\} \cdot \left(
\begin{matrix}
 \mathrm{K}_0(M_{B_{s}} F_{(2)} b_1), & \text{for}\quad F^2_{(2)}>0 \\
 \frac{\pi i}{2} \mathrm{H}_0^{(1)}(M_{B_{s}}\sqrt{|F^2_{(2)}|}\ b_1), &
 \text{for}\quad F^2_{(2)}<0
\end{matrix}\right),
\label{eq:pp3}
 \eeq
 \beq
 h_f^2(x_1,x_2,x_3,b_1,b_2) &=&
\biggl\{\theta(b_1-b_2) \mathrm{K}_0(i \sqrt{x_2 x_3} b_1 M_{B_{s}})
 \mathrm{I}_0(i \sqrt{x_2 x_3} b_2 M_{B_{s}})
  \non
 &+& (b_1 \leftrightarrow b_2) \biggr\} \cdot \left(
 \begin{matrix}
 \mathrm{K}_0(M_{B_{s}} F_{(3)} b_1), & \text{for}\quad F^2_{(3)}>0 \\
 \frac{\pi i}{2} \mathrm{H}_0^{(1)}(M_{B_{s}}\sqrt{|F^2_{(3)}|}\ b_1), &
 \text{for}\quad F^2_{(3)}<0
 \end{matrix}
 \right),
 \label{eq:pp4}
 \eeq
where $J_0$ is the Bessel function,   $K_0$ and  $I_0$ are modified
Bessel functions $K_0 (-i x) = -(\pi/2) Y_0 (x) + i (\pi/2) J_0
(x)$; $\mathrm{H}_0^{(1)}(z)$ is the Hankel function
,$\mathrm{H}_0^{(1)}(z) = \mathrm{J}_0(z) + i\, \mathrm{Y}_0(z)$,
and $F_{(j)}$'s are defined by
 \beq
F^2_{(1)}&=&(x_1 -x_2) x_3\;,\non
F^2_{(2)}&=&(x_1-x_2) x_3\;,\non
F^2_{(3)}&=& x_1+x_2+x_3-x_1 x_3-x_2 x_3 \;\;.
 \eeq

The threshold resummation form factor $S_t(x_i)$ is adopted from
Ref.\cite{tk07}
  \beq S_t(x)=\frac{2^{1+2c} \Gamma
(3/2+c)}{\sqrt{\pi} \Gamma(1+c)}[x(1-x)]^c,
   \eeq where the
parameter $c=0.3$. This function is normalized to unity. More
information about the threshold resummation can be found in
reference \cite{hnl10}.

The Sudakov factors used in the text are defined as
 \beq
S_{ab}(t) &=& s\left(x_1 m_{B_{s}}/\sqrt{2}, b_1\right) +s\left(x_3
m_{B_{s}}/\sqrt{2}, b_3\right) +s\left((1-x_3) m_{B_{s}}/\sqrt{2},
b_3\right) \non
&&-\frac{1}{\beta_1}\left[\ln\frac{\ln(t/\Lambda)}{-\ln(b_1\Lambda)}
+\ln\frac{\ln(t/\Lambda)}{-\ln(b_3\Lambda)}\right],
\label{wp}\\
S_{cd}(t) &=& s\left(x_1 m_{B_{s}}/\sqrt{2}, b_1\right)
 +s\left(x_2 m_{B_{s}}/\sqrt{2}, b_2\right)
+s\left((1-x_2) m_{B_{s}}/\sqrt{2}, b_2\right) \non
 && +s\left(x_3
m_{B_{s}}/\sqrt{2}, b_1\right) +s\left((1-x_3) m_{B_{s}}/\sqrt{2},
b_1\right) \non
 & &-\frac{1}{\beta_1}\left[2
\ln\frac{\ln(t/\Lambda)}{-\ln(b_1\Lambda)}
+\ln\frac{\ln(t/\Lambda)}{-\ln(b_2\Lambda)}\right],
\label{Sc}\\
S_{ef}(t) &=& s\left(x_1 m_{B_{s}}/\sqrt{2}, b_1\right)
 +s\left(x_2 m_{B_{s}}/\sqrt{2}, b_2\right)
+s\left((1-x_2) m_{B_{s}}/\sqrt{2}, b_2\right) \non
 && +s\left(x_3
m_{B_{s}}/\sqrt{2}, b_2\right) +s\left((1-x_3) m_{B_{s}}/\sqrt{2},
b_2\right) \non
 &&-\frac{1}{\beta_1}\left[\ln\frac{\ln(t/\Lambda)}{-\ln(b_1\Lambda)}
+2\ln\frac{\ln(t/\Lambda)}{-\ln(b_2\Lambda)}\right],
\label{Se}\\
S_{gh}(t) &=& s\left(x_2 m_{B_{s}}/\sqrt{2}, b_2\right)
 +s\left(x_3 m_{B_{s}}/\sqrt{2}, b_3\right)
+s\left((1-x_2) m_{B_{s}}/\sqrt{2}, b_2\right) \non
 &+& s\left((1-x_3)
m_{B_{s}}/\sqrt{2}, b_3\right)
-\frac{1}{\beta_1}\left[\ln\frac{\ln(t/\Lambda)}{-\ln(b_3\Lambda)}
+\ln\frac{\ln(t/\Lambda)}{-\ln(b_2\Lambda)}\right], \label{ww}
 \eeq
where the function $S_{B_s}$, $S_{\rho(\omega)}$, $S_{K}$ used in
the amplitudes are defined as:
 \beq
 S_{B_s}(t) &=& s(x_1P_1^+,b_1)+2\int_{1/b_1}^t\!\!\!\frac{d\bar\mu}{\bar\mu}
 \gamma(\alpha_s(\bar\mu)),\\
  S_{\rho(\omega)}(t) &=& s(x_2P_2^+,b_2)+s\left((1-x_2)P_2^+,b_2\right)+
 2\int_{1/b_2}^t\!\!\!\frac{d\bar\mu}{\bar\mu}\gamma\left(\alpha_s(\bar\mu)\right),\\
 S_K(t) &=& s(x_3P_3^-,b_3)+s\left((1-x_3)P_3^-,b_3\right)+
 2\int_{1/b_3}^t\!\!\!\frac{d\bar\mu}{\bar\mu}\gamma\left(\alpha_s(\bar\mu)\right).
 \eeq
where the so called Sudakov factor $s(Q,b)$ resulting from the
resummation of double logarithms is given as \cite{hnl97}:
 \beq
s(Q,b)=\int_{1/b}^Q\!\!\! \frac{d\mu}{\mu}\Bigl[
\ln\left(\frac{Q}{\mu}\right)A(\alpha(\bar\mu))+B(\alpha_s(\bar\mu))
\Bigr] \label{su1}
 \eeq
with
\begin{gather}
A=C_F\frac{\alpha_s}{\pi}+\left[\frac{67}{9}-\frac{\pi^2}{3}-\frac{10}{27}n_{f}+
\frac{2}{3}\beta_0\ln\left(\frac{e^{\gamma_E}}{2}\right)\right]
 \left(\frac{\alpha_s}{\pi}\right)^2 ,\non
B=\frac{2}{3}\frac{\alpha_s}{\pi}\ln\left(\frac{e^{2\gamma_{E}-1}}{2}\right).
\hspace{6cm}
\end{gather}
Here $\gamma_E=0.57722\cdots$ is the Euler constant, $n_{f}$ is the
active quark flavor number.

 The hard scale $t_i$'s in the above equations are chosen as:
 \beq
t_{e}^1 &=& {\rm max}(\sqrt{x_3} m_{B_{s}},1/b_1,1/b_3)\;,\non
t_{e}^2 &=& {\rm max}(\sqrt{x_1}m_{B_{s}},1/b_1,1/b_3)\;,\non
t_{e}^3 &=& {\rm max}(\sqrt{x_3}m_{B_{s}},1/b_2,1/b_3)\;,\non
t_{e}^4 &=& {\rm max}(\sqrt{x_2}m_{B_{s}},1/b_2,1/b_3)\;,\non
t_{f} &=& {\rm max}(\sqrt{x_1 x_3}m_{B_{s}},
\sqrt{x_2 x_3} m_{B_{s}},1/b_1,1/b_2)\;,\non
t_{f}^1 &=& {\rm max}(\sqrt{x_2x_3} m_{B_{s}},1/b_1,1/b_2)\;,\non
 t_{f}^2 &=& {\rm max}(\sqrt{x_1+x_2+x_3-x_1 x_3-x_2 x_3}m_{B_{s}},
  \sqrt{x_2 x_3} m_{B_{s}},1/b_1,1/b_2)\;.
  \eeq
They are given as the maximum energy scale appearing in each diagram
to kill the large logarithmic radiative corrections.

\end{appendix}

%%%%%%%%%%%%%%%%%%%%%%%%%%%%%%%%%%%%%%%%%%%%%%%%%%%%%%%%%%%%%%%%%%%%%%%%%%%%%%%%%%%%%%%%%%%%%%5
%                                 reference
%%%%%%%%%%%%%%%%%%%%%%%%%%%%%%%%%%%%%%%%%%%%%%%%%%%%%%%%%%%%%%%%%%%%%%%%%%%%%%%%%%%%%%%%%%%%%%%%%

%\newpage

\end{document}